# Facade Inspection: Design, Prototyping, and Testing of a Hybrid Cable-Driven Parallel Robot

Ginna Marcela García-Rodríguez[1], Eduardo Castillo-Castañeda[1], Giuseppe Carbone[2], Antonio P. Paglia[2], Manuel Tripodi[2], Med Amine Laribi[3], Abdelbadia Chaker[3]

[1] CICATA Unidad Querétaro, Instituto Politécnico Nacional, 76090, México.
[2] DIMEG, University of Calabria, 87036, Rende, CS, Italy.
[3] Department GMSC–Pprime Institute, CNRS–University of Poitiers–ENSMA, 86073, France.
ginnagarcia89@hotmail.com

**Abstract.** In the field of architecture, early detection of damage or degradation of building facades has become increasingly vital due to the need for continuous monitoring of structural integrity. Traditional methods, such as visual inspections, are being supplemented by technological advancements, especially in robotics, which offer innovative solutions for more efficient and precise inspections. This work focuses on the development of a five degree of freedom hybrid cable-driven parallel robot designed for vertical facade inspections. A detailed robot's design and CAD modeling, with a particular focus on a torque transmission mechanism that simplifies the motion of two cables using a single motor is presented. Two degrees of freedom are driven by cables, while the remaining three are driven by a Sarrus-type mechanism and a pan-tilt mechanism. The inverse kinematics models are also developed. A prototype is presented, involving additive manufacturing. A control system for tracking a zig-zag trajectory, commonly used in inspection tasks, was experimentally validated.

**Keywords:** Cable-Driven Parallel Robot, Facade Inspection, Structural Stability, Trajectory Tracking.

## 1 Introduction

Current infrastructure inspection management methods rely heavily on visual inspection to identify signs of damage on the surface [1]. Various types of structures, including bridges, pipelines, rails and aircraft, are subject to periodic inspections by trained specialists. Although this conventional inspection method simplifies the operational development of the activity, it has also been characterized as a strategy with high levels of risk and costs. Visual inspections are qualitative in nature due to the human subjectivity inevitably introduced into the process. When damage is detected in a structure, quantitative assessment methods are available on the market, such as non-destructive evaluation (NDE) technologies, which include acoustic emissions, ultrasonic inspection, thermography, and ground-penetrating radar imaging [2-7]. These monitoring systems may include permanent sensors installed in buildings to measure structural responses to environmental and operational loads. Due to their high installation and maintenance costs, these types of systems are only installed in critical structures exposed to extreme loading environments. The integration of robotics in the construction industry brings substantial advances, such as a notable increase in productivity, a decrease in errors, risks and costs [8]. However, industrial robots in service applications, such as building inspection, is not feasible due to their rigid design, high adaptation costs, low mobility, energy consumption and lack of adequate safety mechanisms. Instead, our proposal is to design robots for these tasks, with greater autonomy, adaptability and safety. Cable-driven robots represent an alternative solution for inspection tasks due to the characteristics of parallel robots, where rigid links are replaced by cables [9],[10]. The properties of cables, such as their low inertia and wide range of movement, allow their implementation in multiple fields. Then, a Hybrid Cable-Driven Parallel Robot (HCDPR) is proposed as a solution, it presents advantages such as a large translation workspace and light moving parts.

# 2 Robot's design

## 2.1 Robot frame and transmission system

The main frame of the HCDPR has two main planes, see Fig. 1a. A mobile platform is driven by eight cables. A double pulley with a cross belt as is used as a transmission system, see Fig. 1b. A drive pulley is coupled to the motor, the crossed belt causes the opposite rotation of the driven pulley.

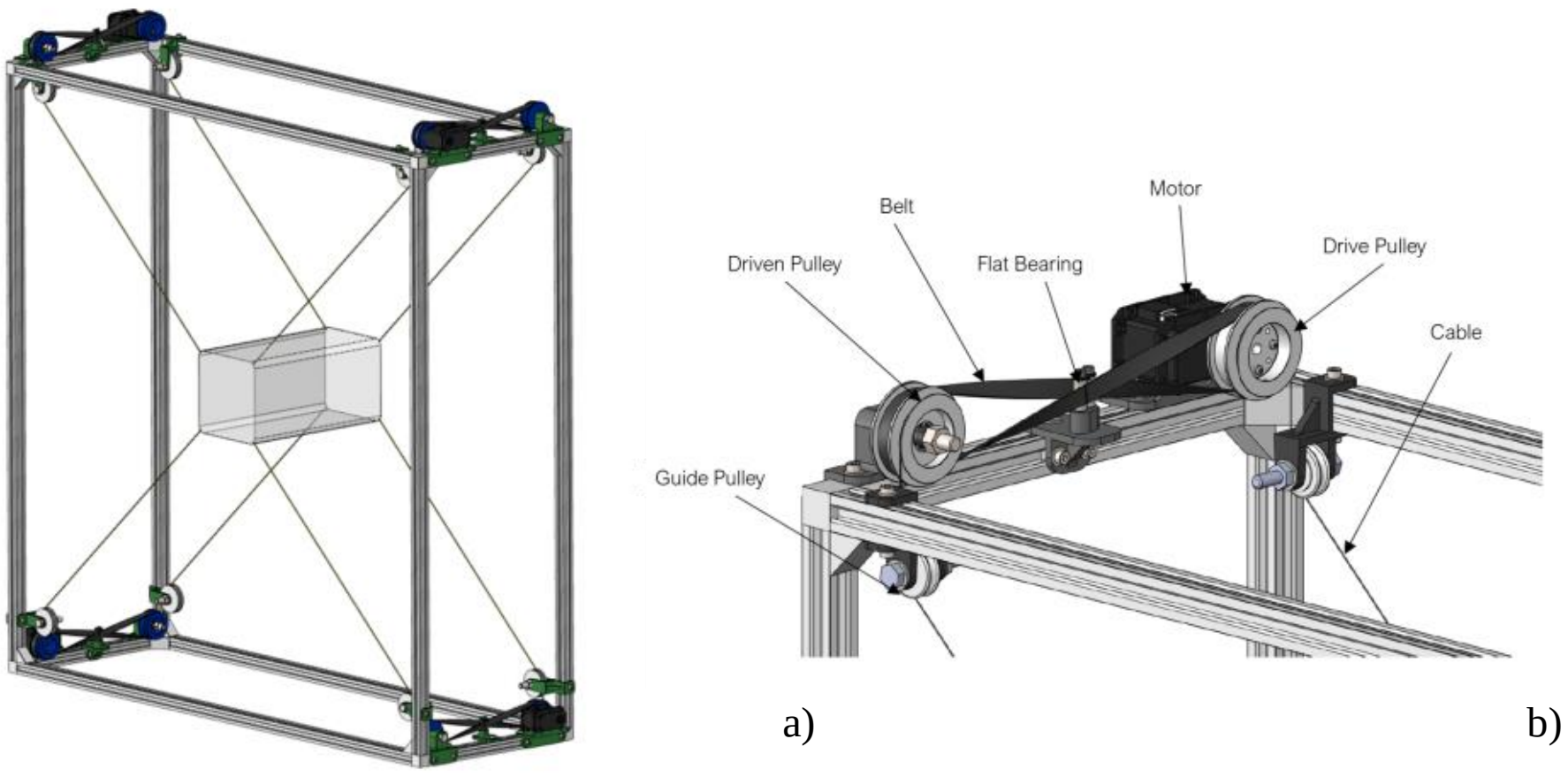

**Fig. 1.** HCDPR robot: a) main frame and mobile platform, b) transmission system.

The HCDPR has 5DOF, two are driven by cables, and the remaining three are driven by a Sarrus-type mechanism and a Pan-Tilt-type mechanism. The requirement focus on the reach of all points on the surface of the facade considering the different profiles (concavities) and the tracking of the required inspection path. A flat cross belt was selected, see Fig. 2, which is most suitable for applications with low speeds and offers an efficiency of 98%, like gears, in addition to generating little noise and absorbing more vibrations than trapezoidal belts or gears.

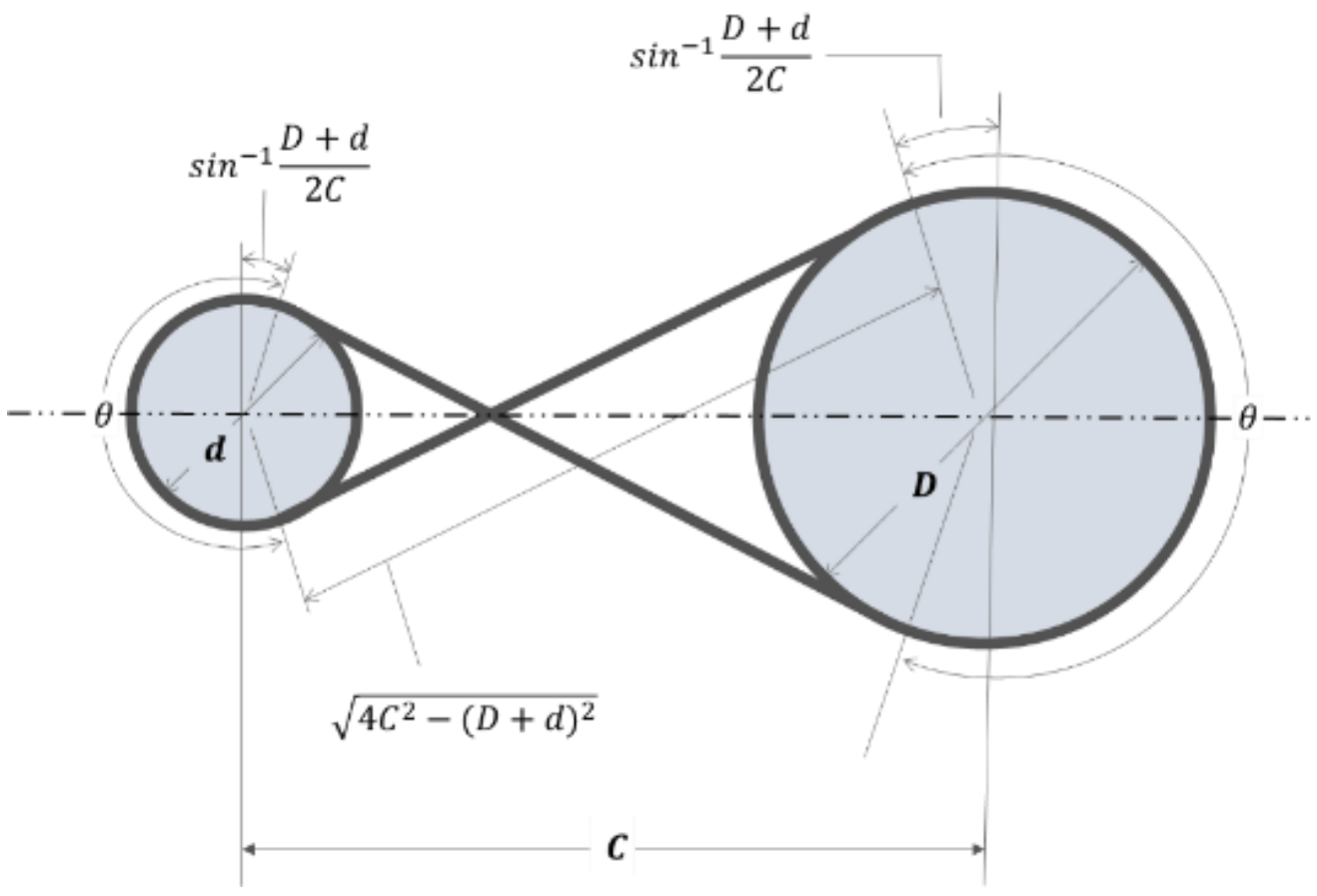

**Fig. 2.** Cross belt transmission.

The contact angle θ is the same for both pulleys and can be calculated using Eq. 1.

$$\boldsymbol{\theta = \pi + 2sin^{-1}\left(\frac{D}{C}\right)} \tag{1}$$

Where:

$\boldsymbol{\theta}$: Contact angle of the belt around the pulley.

$\boldsymbol{C}$: Distance between the centers of the pulleys.

$\boldsymbol{D}$: Diameter of the pulleys (considering both to be of equal dimension).

From Eq. 1 the length of the belt can be obtained, in the case of cross transmission with pulleys of equal diameter.

$$L = \sqrt{4C^2 - 4D^2} + D\theta \tag{2}$$

The first part of Eq.2 refers to the contribution of the linear section of the belt, while the second part refers to the arc that surrounds the pulleys.

## 2.2 Mobile platform design

A mobile platform was designed for the HCDPR, using two acrylic plates connected by aluminum bars to ensure stability and lightness. To determine the dimensions and arrangement of the bars, FEM simulations were performed using Solidworks soft-ware, see Fig.3a. Fixing conditions were applied at the cable attachment points and loads, such as the weight of the end effector, were simulated, resulting in a maximum deformation of 0.57 mm, considered acceptable. The aluminum bars have a diameter of 14 mm and a length of 264.4 mm. The mobile platform also incorporates a component for mounting sensors and an Arduino Uno control board, see Fig. 3b.

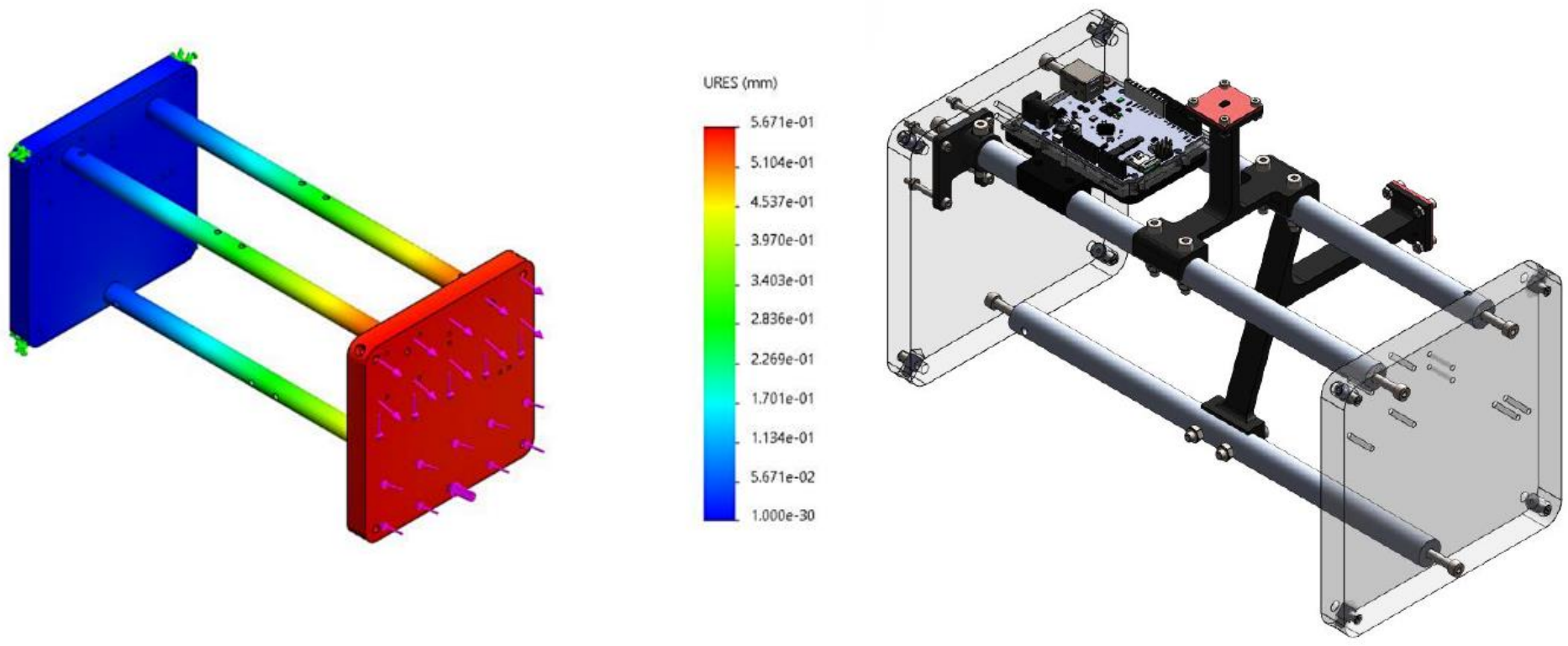

a) b)

**Fig. 3.** Mobile platform design: a) FEM simulations, b) CAD mobile platform.

## 3 HCDPR inverse kinematics

The end effector X-Y position is controlled by four driven pulleys with centers $P_1$, $P_2$, $P_3$ and $P_4$, Fig. 4a. The points $B_1$, $B_2$, $B_3$ and $B_4$, linked to mobile platform are described with respect to a local reference (acrylic plate).

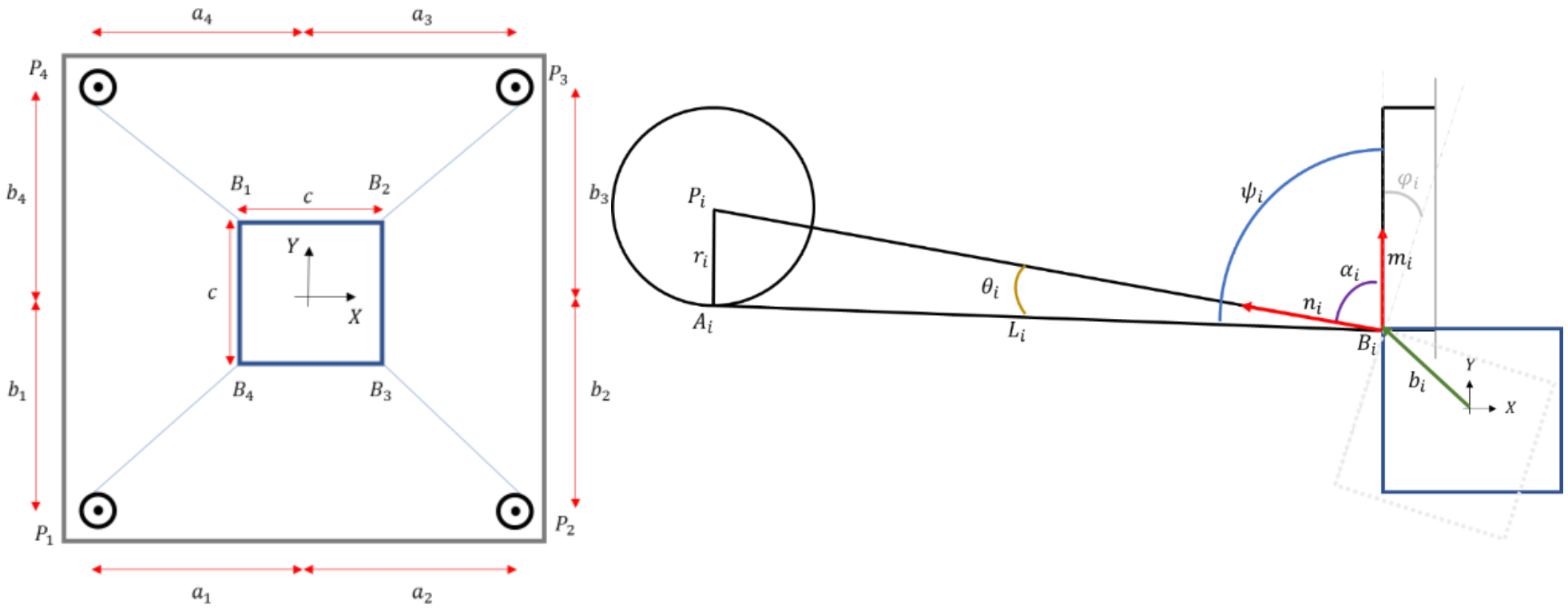

a) b)

**Fig. 4.** a) HCDPR scheme, b) Pulley and cable scheme.

The coordinates of pulley centers are given by Eq. 3:

$$\begin{bmatrix} P_1 \\ P_2 \\ P_3 \\ P_4 \end{bmatrix} = \begin{bmatrix} -a_1 & -b_1 \\ +a_2 & -b_2 \\ +a_3 & +b_3 \\ -a_4 & +b_4 \end{bmatrix} \tag{3}$$

Where $a_i$ and $b_i$ lengths are described in Fig. 4a. The points $B_i$, where the cables join the mobile platform are given by Eq. 4:

$$\mathbf{B} = \begin{bmatrix} B_1 \\ B_2 \\ B_3 \\ B_4 \end{bmatrix} = \frac{1}{2}\begin{bmatrix} -c & +c \\ +c & +c \\ +c & -c \\ -c & -c \end{bmatrix} = \frac{c}{2}\begin{bmatrix} -1 & +1 \\ +1 & +1 \\ +1 & -1 \\ -1 & -1 \end{bmatrix} \tag{4}$$

According to Fig. 4b, the points $A_i$ represent the exit points of the cables, where *i = 1,2,3,4* correspond to the front face of the structure; and where *i = 5,6,7,8* correspond to the back face. The positions of these points are variable and depend on the poses adopted by the end effector. The length of the cable $i$ is denoted by the distance $L_i$, measured between $A_i$ and $B_i$. This distance is calculated from Eq. 5:

$$L_i{}^2 + r_i{}^2 = |P_i B_i|^2 \tag{5}$$

Where $r_i$ represents the radius of the pulleys, it is considered constant, since the diameter of the cables is insignificant and therefore it is neglected. Solving the distance $L_i$ from Eq. 6, we obtain Eq. 6:

$$L_i = \sqrt{|P_i B_i|^2 - r_i^2} \tag{6}$$

Also, $\theta_i$ can be calculated once the cable length is known as expressed in Eq. 7:

$$\theta_i = \operatorname{atan}\left(\frac{r_i}{L_i}\right) \tag{7}$$

In this way, the unit vectors from the points $B_i$ to $A_i$ are obtained, using Eq. (8):

$$n_1 = \begin{bmatrix} -\sin(\psi_1 + \theta_1) \\ -\cos(\psi_1 + \theta_1) \end{bmatrix} \quad n_2 = \begin{bmatrix} \sin(\psi_2 - \theta_2) \\ -\cos(\psi_2 - \theta_2) \end{bmatrix}$$
$$n_3 = \begin{bmatrix} \sin(\psi_3 + \theta_3) \\ \cos(\psi_3 + \theta_3) \end{bmatrix} \quad n_4 = \begin{bmatrix} -\sin(\psi_4 - \theta_4) \\ \cos(\psi_4 - \theta_4) \end{bmatrix} \tag{8}$$

Therefore, the coordinates of point $A_i$ can be determined as (Eq. 9):

$$\mathbf{A_i} = \mathbf{B_i} + L_i \mathbf{n_i} \tag{9}$$

The motor's rotations $\beta_i$ to generate a change in the cable length $\Delta L_i$, are given by Ec. 10:

$$\beta_i = \frac{\Delta L_i}{r_{Mi}} \tag{10}$$

## 4 Prototype manufacturing

The main frame is built with V-Slot type aluminum profiles with a section of 20x20 mm, which forms a three-dimensional frame of 1040x1040x340 mm. The HCDPR is driven by Dynamixel MX 106T servomotors, and its supports and other elements such as pulleys, spacers, among others, were printed using PLA. The cable is a braided nylon thread with a diameter of 0.45 mm with a maximum allowed load of 50 kg. The anti-slip system, located at the belt crossing point as shown in Fig. 5, was completely manufactured by 3D printing. It includes molded flat bearings that allow a continuous flow of the belt without friction. The bushings are fixed with ⌀6 Seeger rings. The belt has a total length of 630 mm, a thickness of 0.9 mm and a width of 10 mm, with a metal core that provides high resistance, meeting the project requirements.

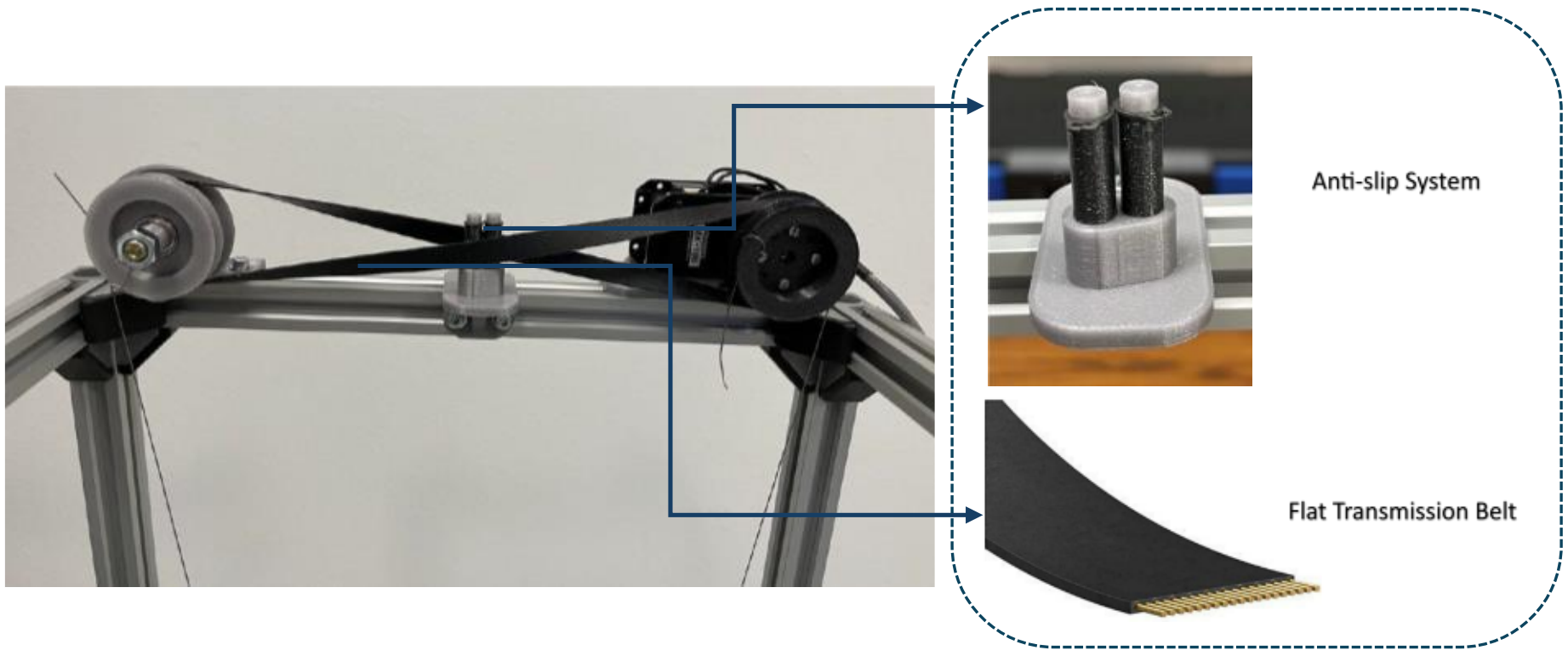

**Fig. 5.** Final prototype of the transmission system.

The subsystem for redirecting the cable to the end effector (Fig. 6) includes guide pulleys with ball bearings to reduce friction. The bracket and spacers were manufactured by 3D printing, the pulley was fixed with a partially threaded M7 screw and nut.

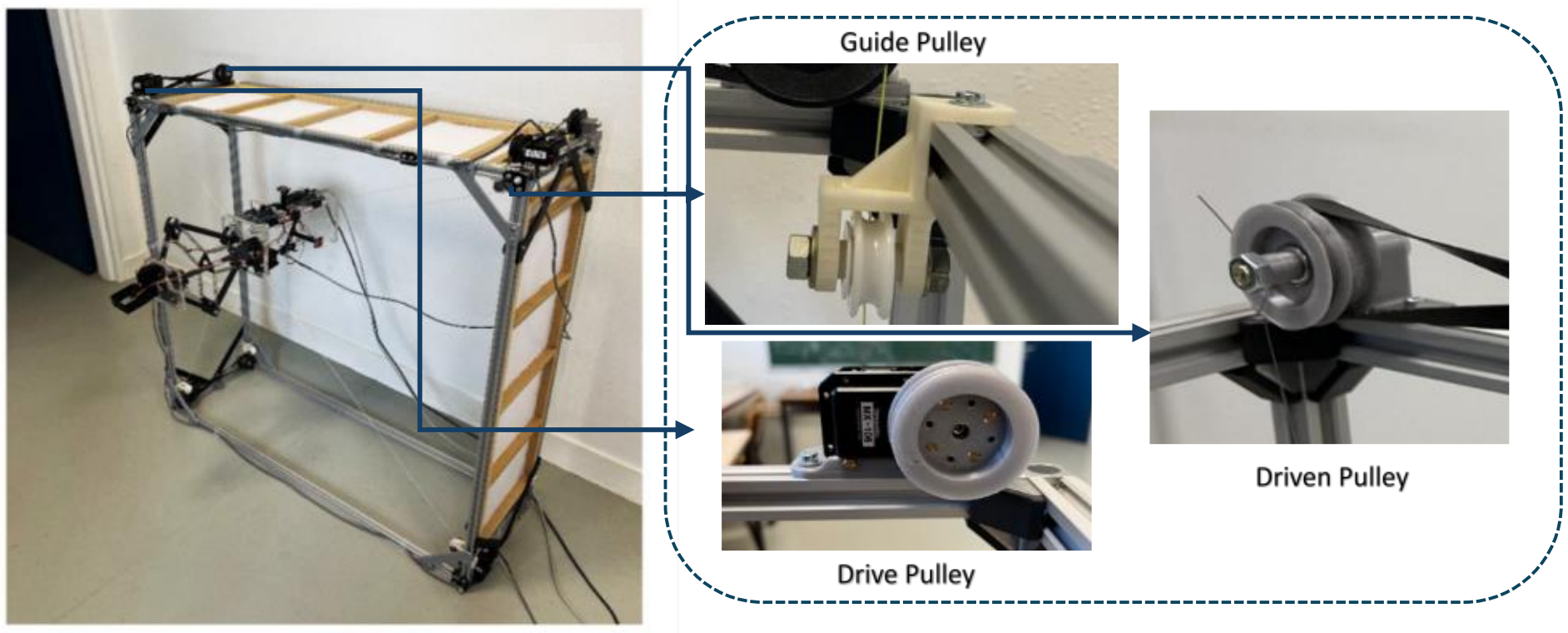

**Fig. 6.** Cable management system.

The mobile platform holds a Sarrus-type mechanism to allow movement in the z axis, see Fig. 7a. The final prototype is shown in Fig. 7b

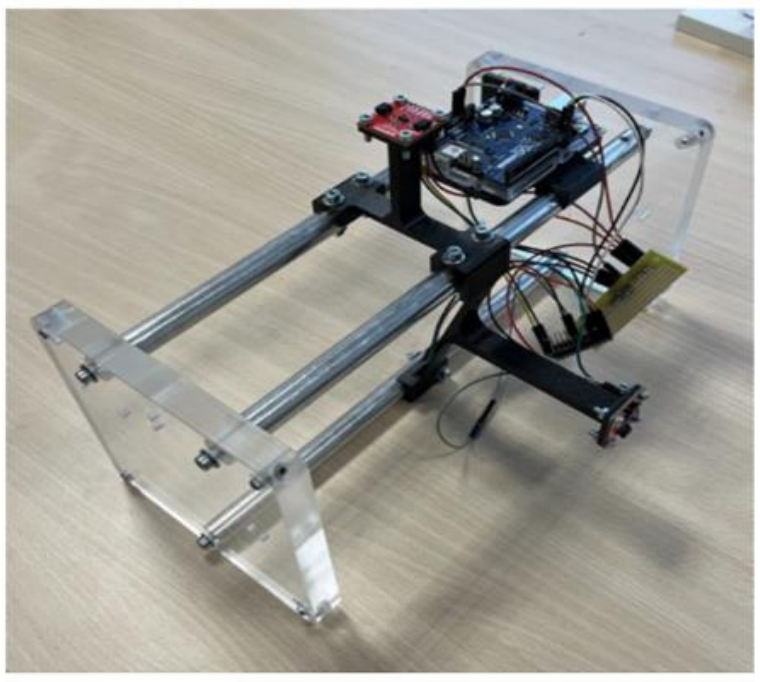

a)

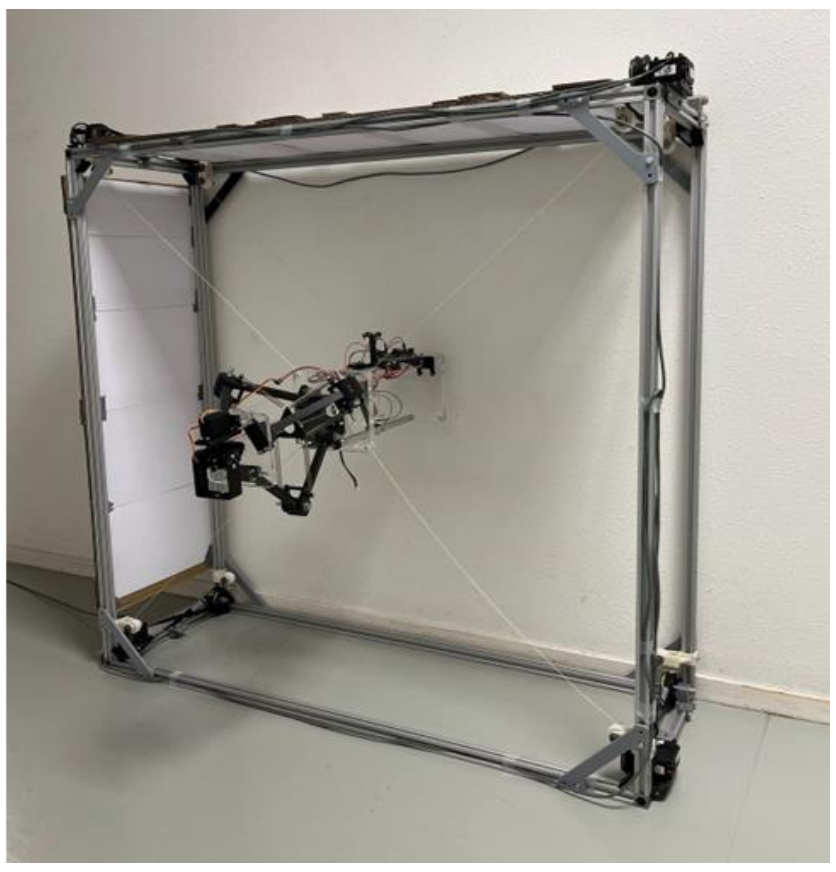

b)

**Fig. 7.** a) Mobile Platform, b) Final prototype.

## 5 Experimental validation

An OptiTrack [11] system collects the data while the robot follows a path. Infrared markers were placed on the main frame and on the mobile platform in such a way that they formed reference systems. A sequence of images representing the movement of the mobile platform is shown in Fig. 8 and then compared with the desired trajectory.

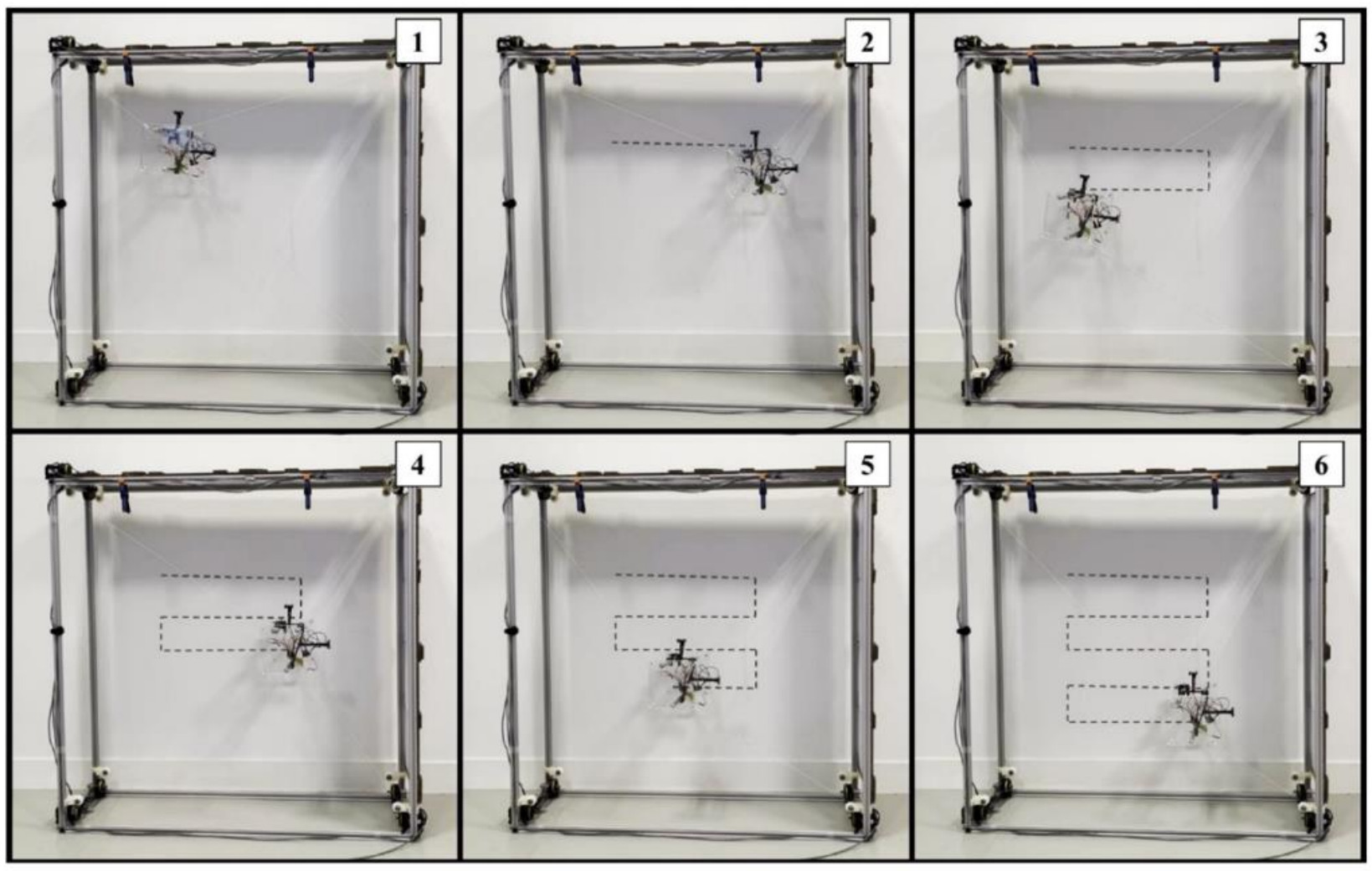

**Fig. 8.** Images sequence during mobile platform motion.

The data obtained from the TOF (Time of Flight) sensor [12] and OptiTrack sensors are comparable, therefore the TOF are suitable for a closed-loop control system, since they are very precise. The real zig-zag trajectory is shown in Fig. 9a, the corresponding motor positions are shown in Fig. 9b.

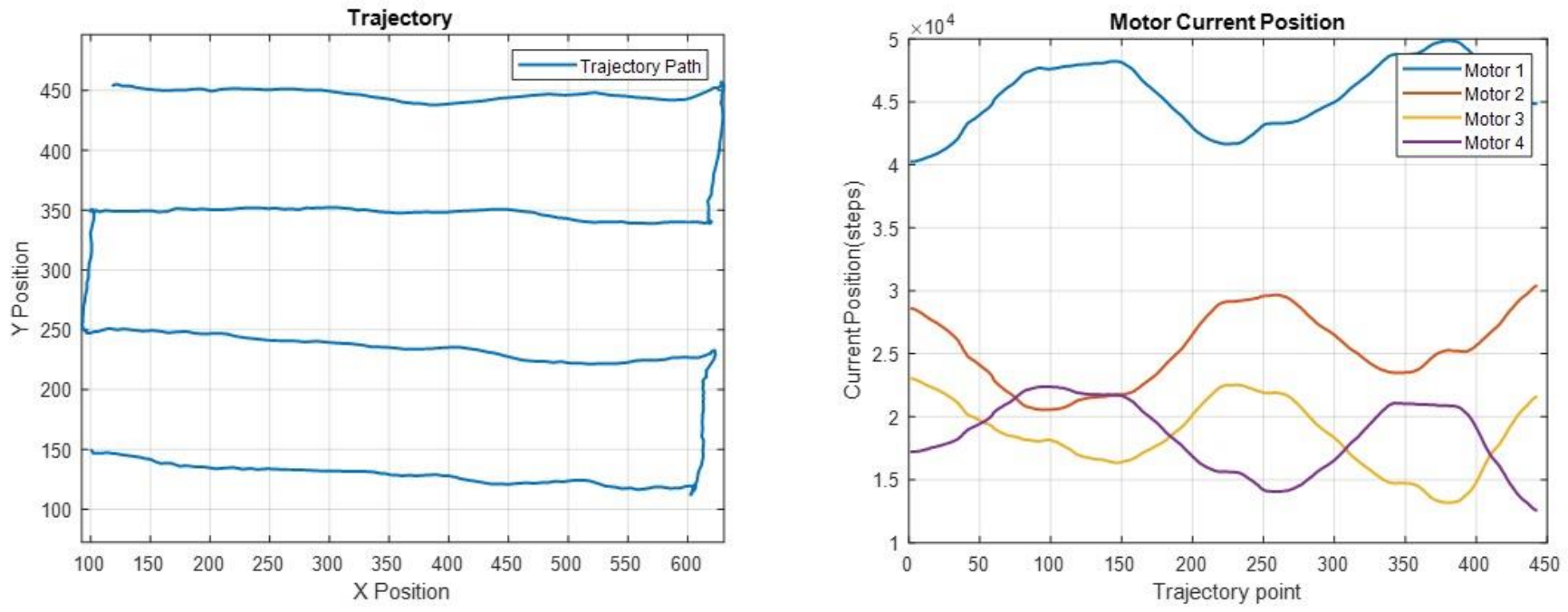

a) b)

**Fig. 9.** a) Real zig-zag trajectory, b) motor's positions.

The evolution of cable length for a Zig-Zag path (Fig. 9), ideal and real, shows that a more precise control is required between the tension and compression elements to achieve the structural principle of tensegrity.

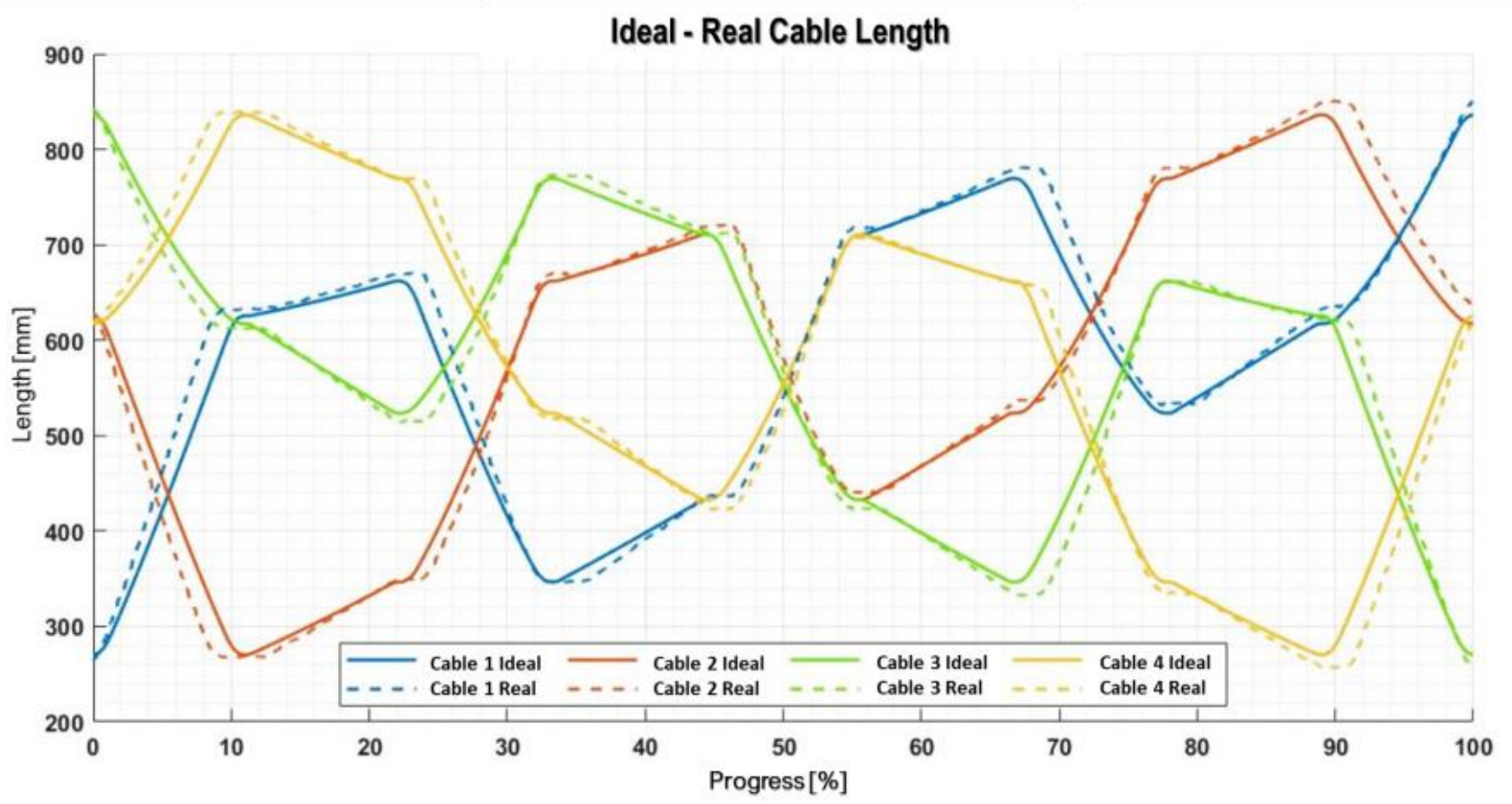

**Fig. 10.** Cable Length Evolution, ideal and real.

Analysis of the graph shows that motors located on the same side, such as Motor 1 and Motor 4, show opposite tendencies due to their counter-rotation. Despite rotating in opposite directions, both motors perform the same function of winding or unwinding the cables.

## Conclusions

The research and development of a hybrid cable-driven parallel robot (CDPR) for vertical facade inspection have demonstrated its feasibility as an effective alternative to traditional inspection methods. The proposed robotic system addresses key challenges associated with visual inspections, such as subjectivity, high costs, and operational risks. By leveraging cable-driven mechanisms, the system benefits from high mobility, low inertia, and scalability, making it suitable for various inspection tasks. The developed prototype integrates a Cartesian and a Pan-Tilt systems to ensure surface coverage. The dual pulley transmission system, driven by a single motor per two cables, enhances movement synchronization while minimizing weight and energy consumption. Additionally, the mobile platform, designed using FEM simulations, proved to be structurally stable, with minimal deformation under operational loads. Experimental validation confirmed the robot’s precise trajectory tracking, with minor deviations due to cable tension variations. The TOF and OptiTrack sensors provided accurate motion data. Future improvements should address tension inconsistencies and vibration reduction to enhance precision. Overall, this work lays the foundation for more autonomous and adaptable robotic inspection solutions, significantly improving safety and efficiency in structural assessments.